\documentclass[11pt]{article}
\usepackage{geometry}                
\usepackage{graphicx}
\usepackage{amssymb}
\usepackage{epstopdf}
\title{A full scale atmospheric flight experimental research environment for the Mars helicopter}
\author{J. Pablo Afman~\footnote{Yamaha Motor Corporation, {\tt juan-pablo\_afman@yamaha-motor.com}} \and Eric Feron~\footnote{Georgia Institute of Technology, {\tt feron@gatech.edu}} \and Mitchell Walker~\footnote{Georgia Institute of Technology, 
{\tt mitchell.walker@ae.gatech.edu}}}

\begin{document}
\maketitle
We propose to develop a full-accuracy flight test environment for the Mars helicopter and related Mars-atmospheric vehicles. The experiment would use reduced-$g$ atmospheric flights with an aircraft that houses a properly sized vacuum chamber.

\section*{Introduction}
Reduced atmospheric density flight has been the object of much interest throughout the history of aviation. Indeed, aviation {\em is} about {reduced air density} flight: 
A jet liner takes off from Boston Logan Airport at sea level. There air density is 1.225 kg/m$^3$. As it crosses the Atlantic ocean, it reaches an altitude up to 38,000 feet above sea level, where air density is approximately reduced to 1/5th of what it is an the ground.  It takes flying up to 100,000 feet above sea level for the atmospheric density to reduce to 1/100th of its sea level density, which corresponds to the atmospheric conditions on Mars' surface. On Earth, several aircraft are capable of flying at 100,000ft. They include the X15 and the Helios solar-powered aircraft. 

NASA and the Jet Propulsion Laboratory (JPL) are in the process of developing a "Mars Helicopter Scout" (MHS) capable of sustained flight over the surface of Mars. This helicopter is the latest of a long sequence of atmosphere-borne candidate Mars vehicles that include Aurora Flight Sciences' ARES, a "Mars airplane" that would be directly dropped from a re-entry vehicle. A complete table of Mars airplane concepts can be found on Wikipedia, \\ see {\tt https://en.wikipedia.org/wiki/Mars\_aircraft}. Earth-based testing of atmospheric Mars vehicles offers good potential to mitigate the possibility for financial and time losses associated with typical Mars missions. So far, the experimental tests performed by JPL include a static flight of the MHS in JPL's 25 ft vacuum chamber with reduced atmospheric density,\\ see {\tt https://www.youtube.com/watch?v=tMCJGfwj3rY}. Other tests include vibration tests and operation at very low temperatures. A core issue is reproducing the Mars gravity conditions. The Jet propulsion Laboratory has addressed this problem by "emulating" the 0.39 $g$ on Mars surface by "assisting" the Mars aircraft with a cable-based gravity reduction mechanism. In the following, we propose an alternative flight testing arrangement that combines the lower Mars gravity with lower density atmospheric effects using a standard cargo aircraft with an embedded vacuum chamber.

	\begin{figure}
		\centering
		\includegraphics[width=.4375\textwidth]{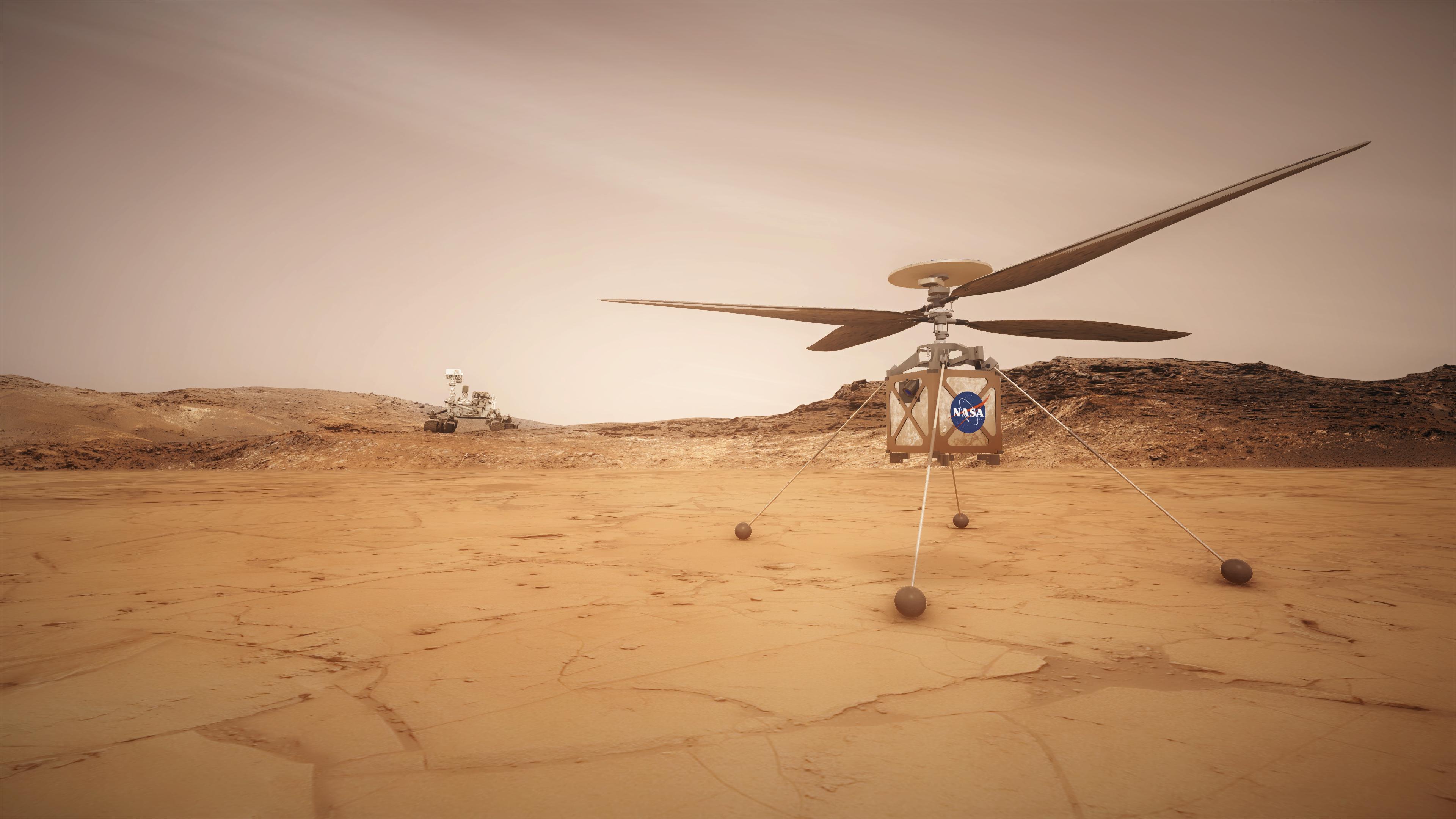}
		\includegraphics[width=.5\textwidth]{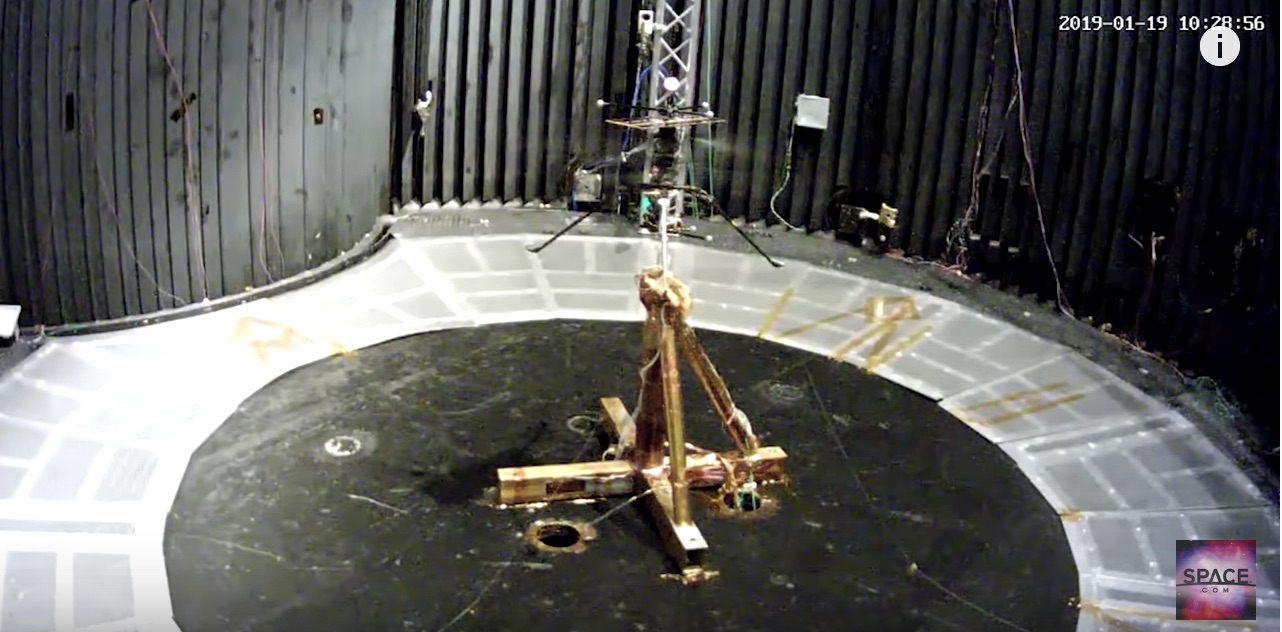}
		\caption{Left: JPL Mars Helicopter Scout (MHS) [?]. Right: JPL Mars Helicopter in JPL's 25ft vacuum test chamber.}
		\label{img:mhs}
	\end{figure}

\section*{Reduced gravity atmospheric flight}

\subsection*{General considerations}
\begin{figure}
		\centering
		\includegraphics[width=.33\textwidth]{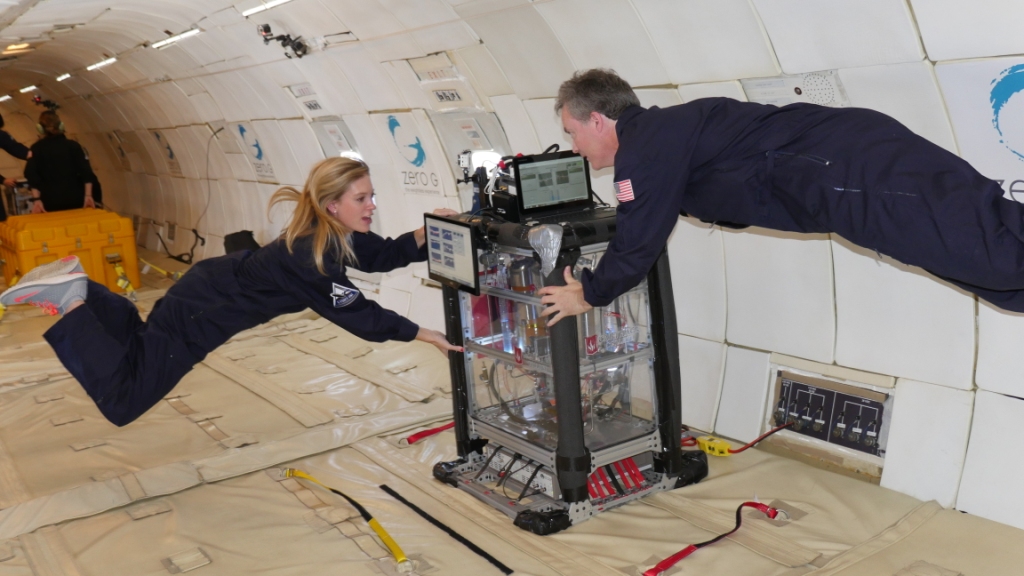}
		\includegraphics[width=.33\textwidth]{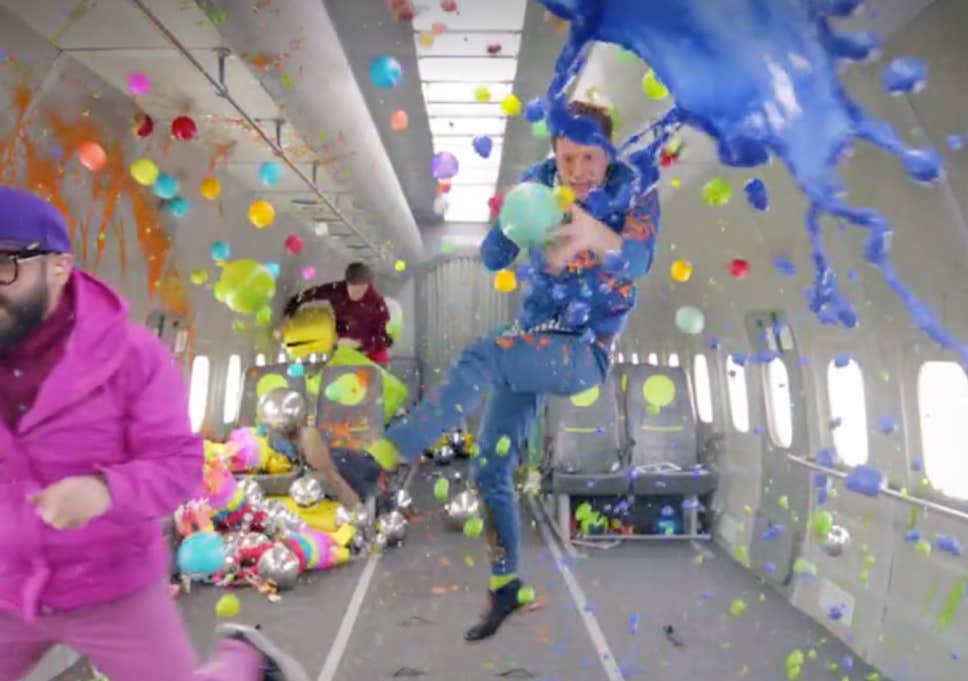}
		\includegraphics[width=.33\textwidth]{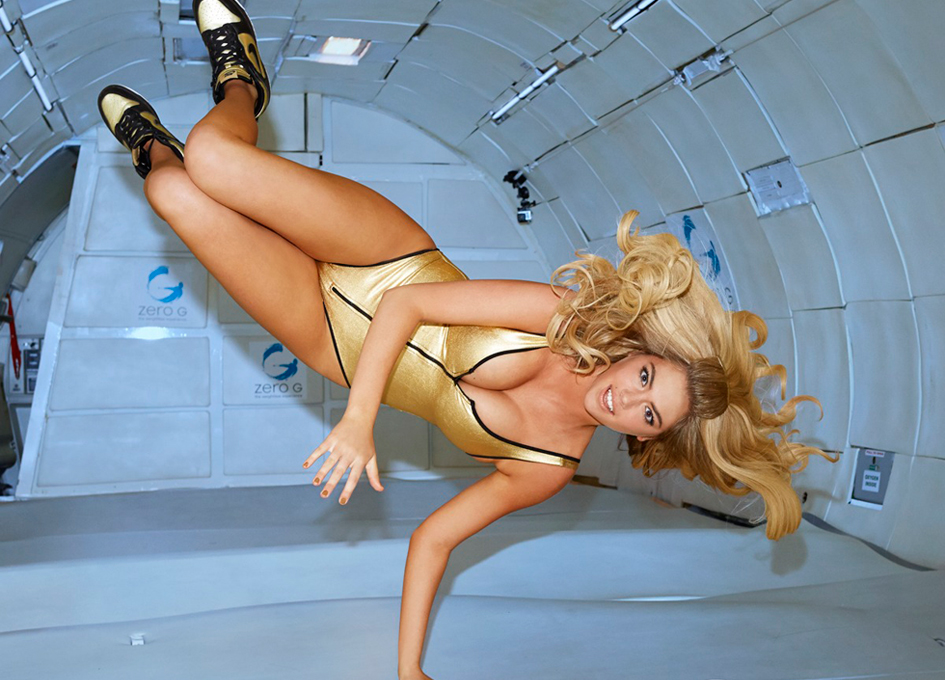}
		\caption{Various contemporary uses of atmospheric zero-$g$ flights. Left: Experimental research. Middle: OK Go, "OK Go - Upside Down \& Inside Out" video. Right: Kate Upton posing in the Zero-$g$ corporation Boeing 727 zero-$g$ aircraft for an advertishttps://www.overleaf.com/project/5d75a44eac3a150001264788ement by SI Swimsuit.}
		\label{img:zero_g_exp}
	\end{figure}

Reduced-gravity atmospheric flight consists of using an atmospheric vehicle (typically an aircraft, but not necessarily, \\ see {\tt https://ieeexplore.ieee.org/document/8431251} and \\ {\tt https://ieeexplore.ieee.org/document/8618690} for example) 
that flies along trajectories where a given level of gravity is "felt" in the reference frame of the vehicle. The most popular form of reduced-gravity flight is the zero-$g$ flight, whereby the aircraft follows exactly the part of an Earth orbit to reproduce weightlessness conditions. In practice actual trajectories, inexactly called parabolic trajectories, last on the order of 18 seconds.  
A variant on zero-$g$ flight is {\em micro}-$g$ or $\mu g$ flights, whereby micro-gravity conditions are created to reproduce those encountered in low-gravity environments, such as asteroids. Accurate $\mu g$ flights are considerably more difficult to create than zero-$g$ flights: the latter can be easily regulated by a skilled pilot by "controlling" the test aircraft against the reference trajectory provided by a proof mass (reportedly a plastic duck initially sitting on the pilot's knees sometimes), and deviations from the nominal trajectories do not matter as long as the proof mass does not deviate exaggeratedly from its free-floating position. Coarsely speaking, that means that human and material subjects in floating conditions will also not move much relative to the aircraft fuselage.  These zero-g tests not only are very popular to achieve purposes of scientific and engineering interest, but also have been used to make video clips for music bands and define innovative environment for fashion shows, see Fig.~\ref{img:zero_g_exp}. In comparison, micro-$g$ is about producing very precise gravity conditions for objects that are {\em fixed} relative to the aircraft. The proof-mass concept then does not work anymore and very precise regulation needs taking place using other sensors than proof masses. 

Creating a Mars gravity environment is similar in nature to the foregoing activities. The point is to create an environment where local gravity is approximately Mars', that is, 0.39 $g$ (or 3.711 m/s$^2$). Flying 0.39 $g$ trajectories is nearly the same as flying a zero-$g$ trajectory, only that the near parabolic trajectory of the aircraft must mimic being pulled to the ground in the vacuum with a constant gravitation of 0.61 $g$ instead of 1 $g$. Such demand on the aircraft is less aggressive than performing a 0-g maneuver, especially during maneuver recovery. \\ According to {\tt https://www.youtube.com/watch?v=tMCJGfwj3rY} , the "useful part" of the MHS flight test is approximately 25-30 seconds, which places the flight within the time window offered by a mars $g$ flight, described below.

\section*{Flight characteristics}

The characteristics of an atmospheric, Mars-$g$ parabolic flight are similar to that of a standard zero-$g$ flight: after a horizontal, rectilinear acceleration phase, the test aircraft initiates a pull-up maneuver so as to reach the proper initial attitude and speed to perform the Mars-$g$ phase of the flight, which resembles an inverted parabola. Once the Mars-$g$ maneuver is complete, the aircraft pulls up to resume straight and level flight. Other maneuvers may follow along the same principle. The challenges that come with the design of these maneuvers include the necessity to avoid stall at all times on the one hand, and to keep Mach number below transonic regime, on the other hand. In addition, we have added the constraint that the maneuver be performed by a standard cargo airliner, rather than a specifically modified aircraft, such as the Zero-g corporation's Boeing 727 or the European Space agency's Airbus A310. The latter aircraft are instrumented with pumps and special equipment that allows the aircraft to operate in zero-g conditions without any issues. For our Martian maneuver, this type of special equipment is likely unnecessary, and a maximum aircraft load limit has been placed at at 1.3 $g$. 

\begin{figure}
		\centering
		\includegraphics[width=.48\textwidth]{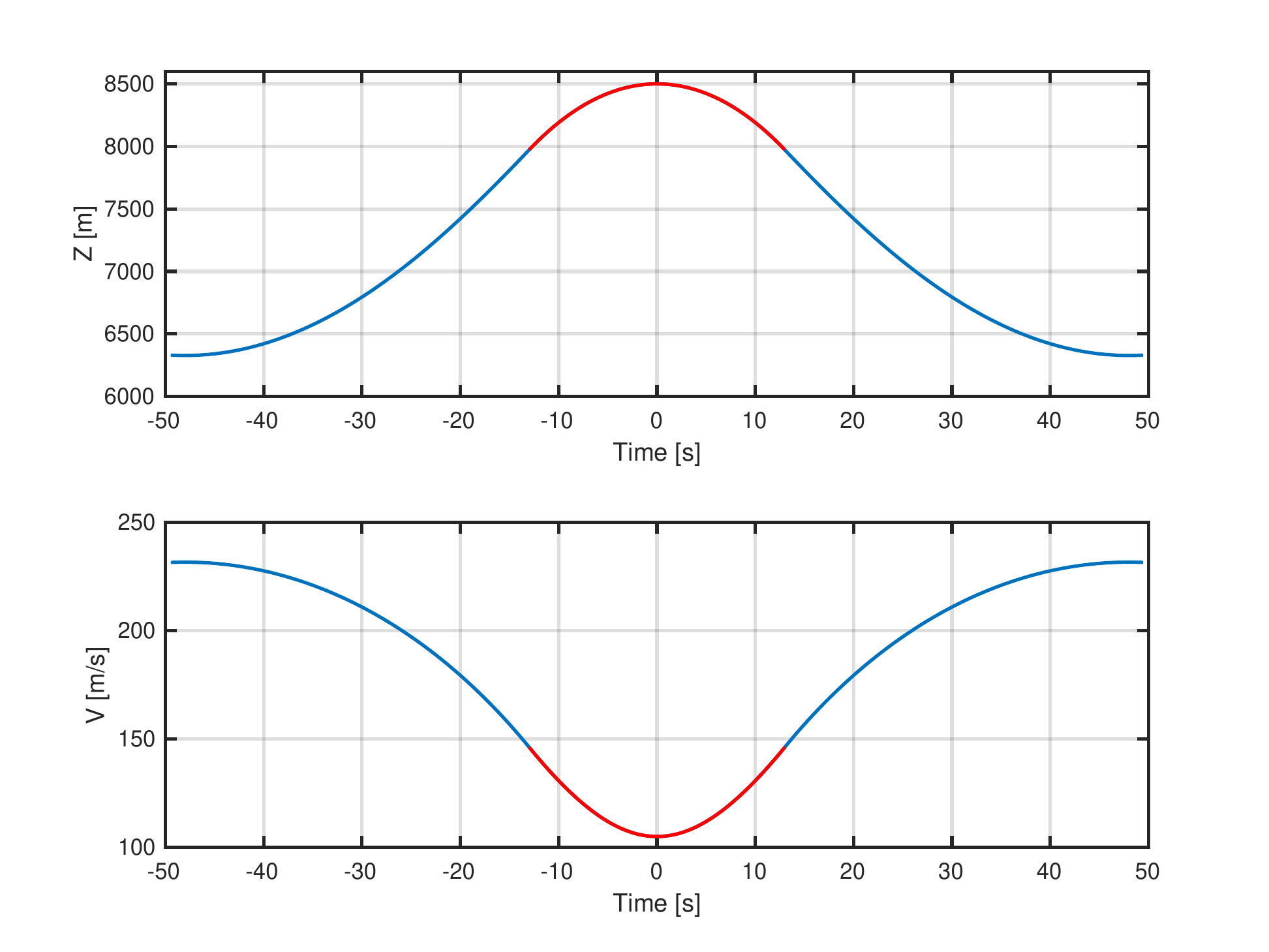}
		\includegraphics[width=.48\textwidth]{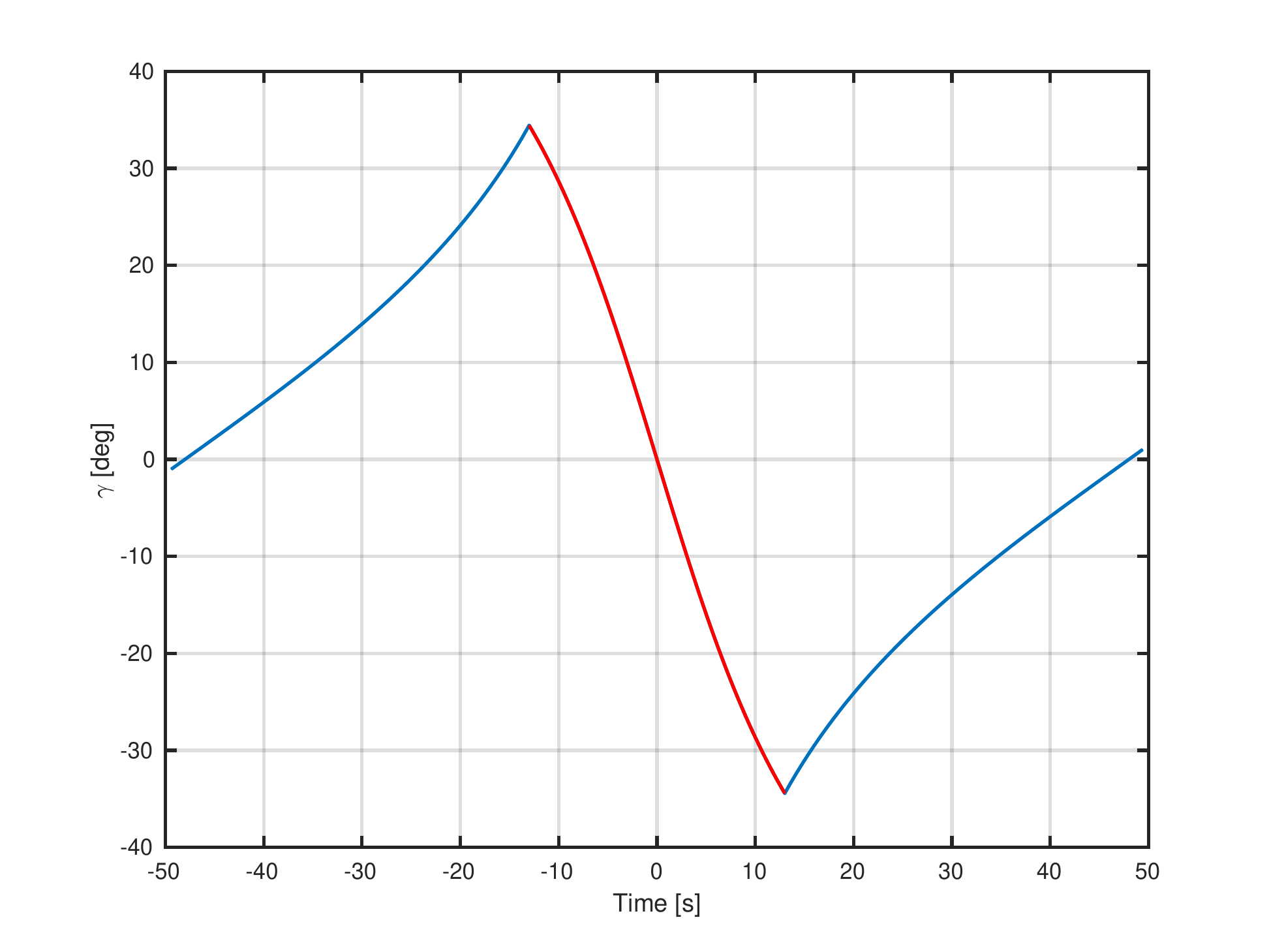}
		\caption{Mars-$g$ trajectories for standard cargo transport aircraft. The Red dashed lines indicate the boundaries of the Mars-$g$ maneuver. Left: aircraft altitude and speed as a function of time (units are second, meters, and meters per second, respectively). Right: Flight path angle.}
		\label{img:mars_g}
\end{figure}
The simulation shown in Fig.~\ref{img:mars_g} indicates that a 26 sec. Mars-$g$ maneuver is achievable without inducing excessive stress on the aircraft (1.3 $g$ max). Moreover, the flight path angle does not exceed 35 degrees in magnitude. On top of the maneuver, true airspeed is 105 m/sec, 
resulting in a maximum angle of attack $< 8$ degrees at  apogee, well within its stall envelope. Such a maneuver may easily be performed by an airline pilot after moderate training to handle upsets or unusual attitudes, \\ see {\tt https://www.faa.gov/regulations\_policies/handbooks\_manuals/aviation/}
\\ {\tt airplane\_handbook/media/06\_afh\_ch4.pdf}.

\section*{Reduced atmospheric density chamber}
While this report does not intend to enter into the details of the vacuum chamber design, several remarks can be made about some of the constraints that must be met before the system can be constructed. 
\subsection*{Geometric constraints}
\begin{figure}[h]
		\centering
				\includegraphics[width=.45\textwidth]{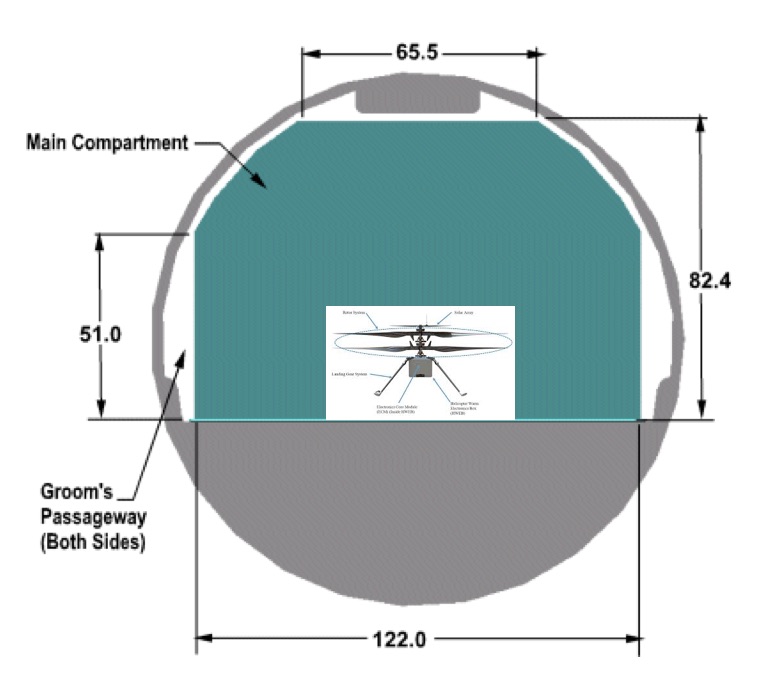}
		\includegraphics[width=.45\textwidth]{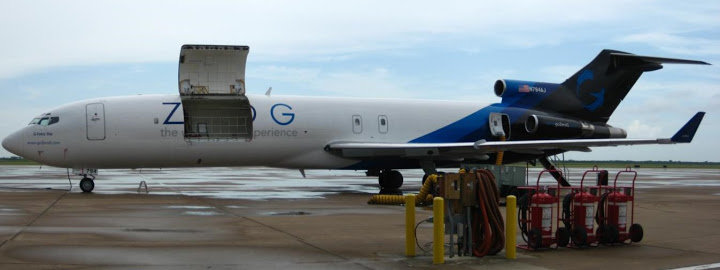}
		\caption{Left: Mars helicopter in B727 aircraft. Relative dimensions are approximate. Right: B727 "vomit comet" zero-$g$ aircraft with large cargo door}
		\label{img:727-helico}
\end{figure}

First, there are the geometric constraints posed by the dimensions of the test aircraft. In short, the bigger the aircraft, the better. 
As a benchmark, the cargo bay of today's reduced $g$ research aircraft in the US, a 727-200 operated by the Zero-$g$ corporation, is about three meters, thus providing a comfortable fit for the "naked" mars helicopter as shown in Fig.~\ref{img:727-helico} ßif it were flying within a standard container  with empty side space at around 1 meter on both sides. Vertical clearance is about  two meters. Container length is limited by cargo door width (about 3.2 meters), as shown in Fig.~\ref{img:727-helico}.

\begin{figure}
		\centering
				\includegraphics[width=.90\textwidth]{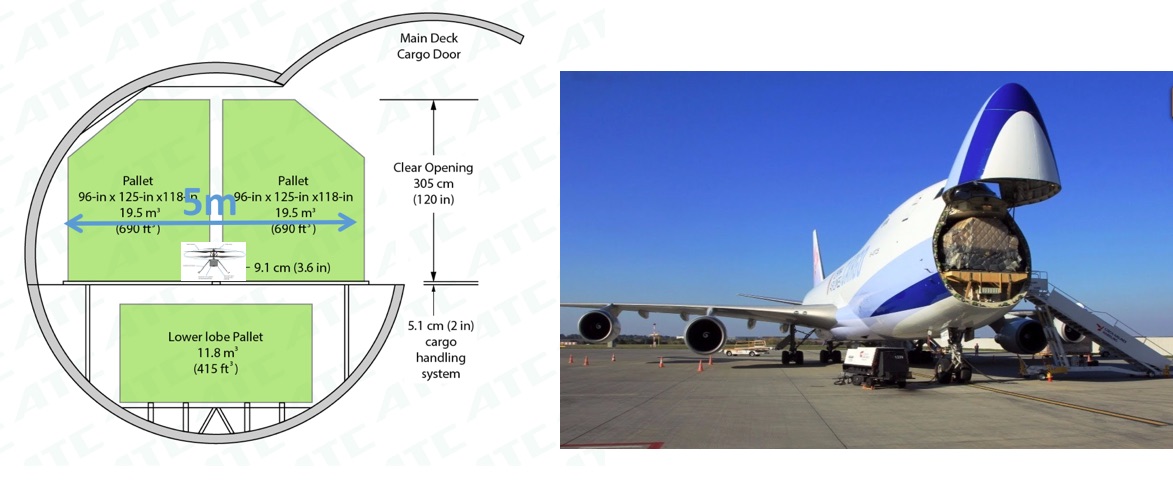}
				\caption{Left: Mars helicopter in B747 cargo aircraft. Relative dimensions are approximate. Right: 747 front cargo door.}
		\label{img:747_helico}
\end{figure}

The European zero-$g$ flight test aircraft, a converted Airbus A310, offers an alternative to the 727. According to the web site describing the characteristics of the payload bay,  see {\tt https://m.esa.int/Our\_Activities/}\\{\tt Human\_and\_Robotic\_Exploration/Research/Airbus\_A310\_Zero-G}, the dimensions of the testing volume are 20 x 5 x 2.3 metres (L x W x H), thus offering superior space available for experiments. However, according to the same web site, the door for equipment loading has a height limit of 1.80 metres and a width limit of 1.06 metres. Thus the vacuum chamber would almost certainly have to be build from several sections and assembled inside the aircraft.
\begin{figure}
		\centering
				\includegraphics[width=.45\textwidth]{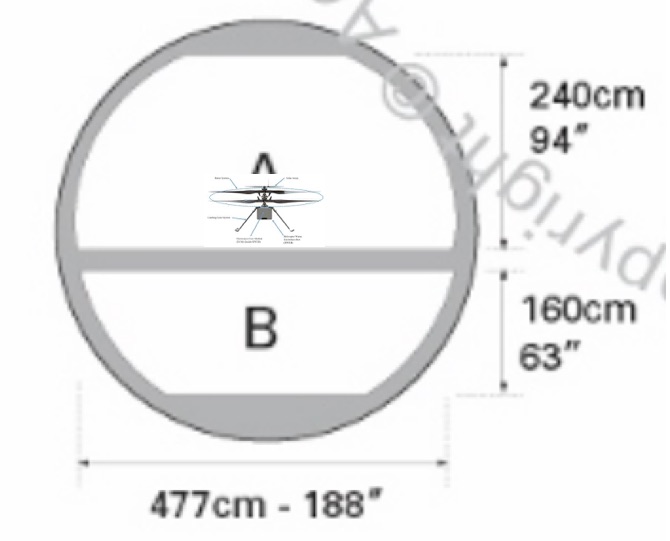}
		\includegraphics[width=.45\textwidth]{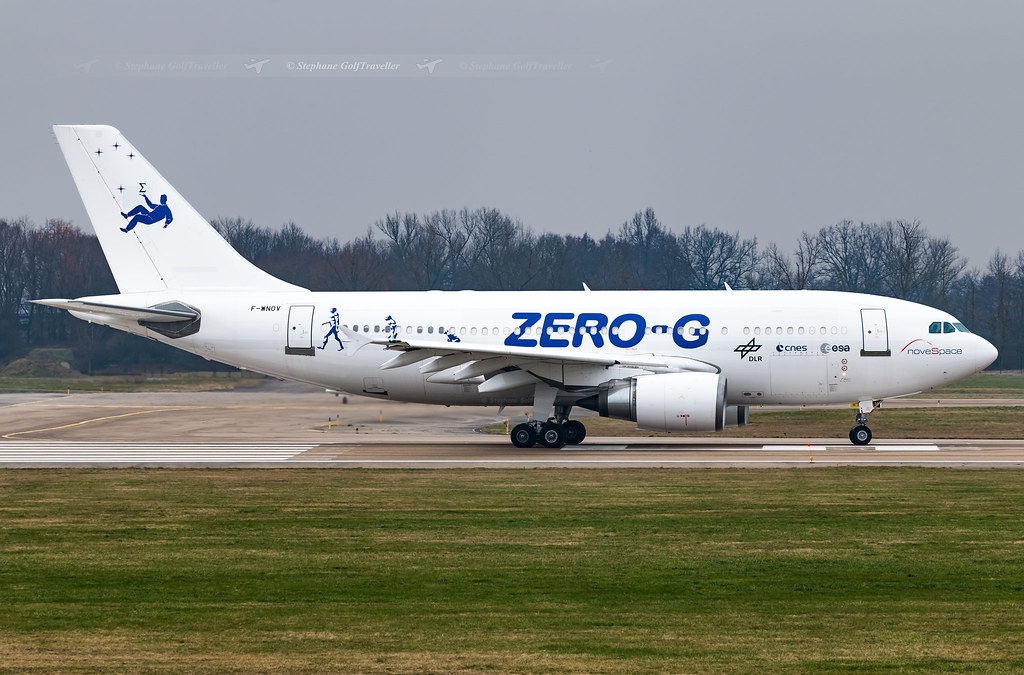}
				\includegraphics[width=.45\textwidth]{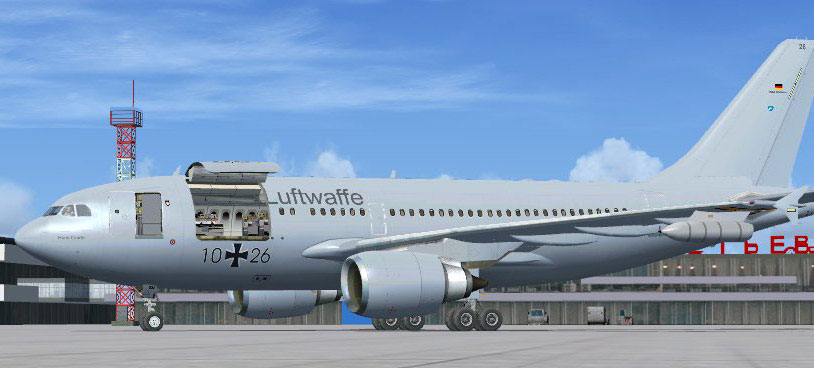}
		\caption{Left: Mars helicopter in A310 aircraft. Relative dimensions are approximate. Right: ESA A310 zero-$g$ aircraft. Doors are those of a standard commercial jet. Bottom: A310 cargo aircraft with large cargo door.}
		\label{img:310_helico}
\end{figure}
Perhaps preferably, a standard A310 cargo aircraft might be used because of the benign flight conditions and the presence of a much larger cargo door, as shown in Fig.~\ref{img:310_helico}. With a relatively low stress on the aircraft, the relatively benign nature of the complete maneuver, and the closed nature of the proposed experiment, it can be surmised that an even larger cargo aircraft can be used to perform the experiment with no additional concern for safety. This opens up the possibility of using considerably larger cargo aircraft, eg one among several available Boeing 747 freighters, whose cargo doors are multiple and very large, as seen on Fig.~\ref{img:747_helico}.
That possibility opens the perspective for using a much larger vacuum chamber whose horizontal dimensions will be close to those used by the Jet Propulsion Laboratory's 25-ft solar thermal vacuum chamber to flight test the current Mars helicopter prototype. 

\subsection*{Aerodynamic constraints}
There are clear challenges of operating a flying machine in a confined space such as a vacuum chamber: managing the gas flows and limiting boundary effects come first. It is well-known that wind-tunnel boundary effects can be deleterious to experimental results. Some of these effects can be mitigated by adding well-located louvers and to recirculate the air appropriately. The larger the vacuum chamber, the more accurate the flight test relative to actual Mars conditions. Preliminary computational fluid dynamics can help lift these uncertainties, just as they can lift those associated with the flight test that took place in JPL's 25-ft vacuum chamber. 

\subsection*{Navigation issues}
A proper, full-scale flight test environment should also be capable of replicating the conditions necessary for the unmanned vehicle to navigate its environment. The MHS navigation system includes a combined INS-vision navigation system. Algorithms used to extract system position and orientation rely on the necessary relations that link GPS readings and optical information in fixed environments. There is a risk, however, that such algorithms might be fooled if the airplane goes through perturbations, such as turbulence. In that case, the relation between optical and inertial readings could be temporarily de-correlated. The question as to whether such perturbations are observable and 
rejected requires more work than that envisioned to prepare the present report.

\section*{Complementarity with other tests}
The core benefit of the proposed test over ground-based tests is the possibility of exactly reproducing the gravity conditions encountered on Mars using an atmospheric device. In addition, it is also possible to effect large attitude changes on the coaxial helicopter, something strictly impossible to do if the machine is suspended to a cable to emulate low gravitational conditions. Moreover, it becomes possible to obtain a better idea of the helicopter behavior as it takes-off {\em from the ground}, including if it takes off not exactly horizontal, a distinct possibility when landing in a largely unknown area, and despite local leveling opportunities offered by robotic ground platforms. Last, it becomes possible to study the very real possibility of "brownout" that may occur when dust gets blown away by the airflow created by the helicopter. Much is known about Mars dust and could be reproduced for the specific purposes sought for the proposed experiment.
\section*{Other proposed uses}
A high-fidelity Mars environment may be used in several ways. For example, there might be value testing Mars probe landing mechanisms, and Mars rovers, at least those whose dimensions are acceptable. There might also be the possibility of testing human response to Mars' gravity environment as part of a human deployment on the planet. Coming back to flying vehicles, the proposed environment may also be used to test smaller systems whose capabilities could eventually match those of the current Mars Helicopter thanks to the rapid evolution and miniaturization of computer, sensor, and actuator hardware, and aerodynamics that scale in favor of smaller systems.

\section*{Acknowledgements}
The authors would like to thank Mr. Olivier Toupet from the Jet Propulsion Laboratory for his insightful comments, most notably for bringing our attention to the navigation issues that arise during the flight test of the Mars helicopter in a confined environment. 
\end{document}